\documentclass{article}

\usepackage{graphicx}
\usepackage[hypertex,dvipdfm,colorlinks=true,urlcolor=blue,linkcolor=blue,citecolor=blue]{hyperref}

\begin{document}
\begin{center}
\textbf{Coexistence of superconductivity and charge-density-wave domain in $1T$-Fe$_x$Ta$_{1-x}$SSe}\\
~\\

Y. Liu, W. J. Lu, L. J. Li, R. Ang, Y. P. Sun\\
Key Laboratory of Materials Physics, Institute of Solid State Physics, Chinese Academy of Sciences, Hefei 230031, People's Republic of China\\
~\\

\end{center}

\textbf{Abstract}
A series of $1T$-Fe$_x$Ta$_{1-x}$SSe (0 $\leq x \leq$ 0.1) single crystals was fabricated via the chemical-vapor-transport (CVT) method and investigated by structure, transport, and magnetic measurements along with the density-functional-theory (DFT) calculations. The superconductivity (SC) in parent $1T$-TaSSe can be gradually suppressed by Fe-substitution ($x\leq0.03$), accompanied by the disappearance of charge-density-wave (CDW). DFT calculations show that the Fe-substitution effectively inhibits the CDW superstructure and thereby the CDW domains are destroyed. With further increasing $x$ ($x>0.03$), the disorder-induced scattering increases, and the system enters into the possible Anderson localization (AL) state. Our results prove the SC develops in the CDW phase and coexists with the CDW domain in $1T$-TaSSe system.\\
~\\
~\\

Superconductivity (SC) often emerges in the vicinity of symmetry-breaking ground state, such as stripe order in cuprates,\cite{Tranquada1995} spin-density-wave (SDW) in iron arsenides,\cite{Canfield2010} and especially, charge-density-wave (CDW) in transition-metal dichalcogenides (TMDs).\cite{Rossnagel2011,DiSalvo1976,PRL2009} The interplay between the electronic order and SC has been extensively studied for decades, however, the understanding that electronic order competes, coexists, or supports SC remains mostly controversial.

The classical CDW is the periodic modulation of conduction electron density in solids, which is generally considered as competing with SC. In cuprates, the competition was early supposed in La$_{1.8}$Ba$_{0.2}$CuO$_4$ on the basis of heat capacity and optical studies,\cite{Gabovich1987} and recently, was directly observed in YBa$_2$Cu$_3$O$_{6.67}$ via high-energy X-ray diffraction (XRD) experiment.\cite{Chang2012} However, in the organic material $\alpha$-(BEDT-TTF)$_2$KHg(SCN)$_4$, the SC coexists with CDW under pressure, which is realized within boundaries between CDW domains,\cite{Kartsovnik2004} similar to the SC in NbSe$_3$ at ambient pressure.\cite{Briggs1981} Recently in $1T$-TiSe$_2$, the high-pressure XRD experiment also indicates that its SC is likely associated with the formation of CDW domain walls.\cite{Joe2013}

Among many TMDs, $1T$-TaS$_2$ has attracted special attention due to the formation of several CDWs. Below $T=$ 180 K, it forms a commensurate CDW (CCDW) order with $\sqrt{13}\times\sqrt{13}$ superstructure.\cite{Wilson1975,Fazekas1979,Scruby1975,Whangbo1992} With increasing temperature, it melts into a nearly-commensurate CDW (NCCDW), which forms roughly hexagonal CDW domains.\cite{Spijkerman1997} The electronic phase diagrams in pressurized $1T$-TaS$_2$,\cite{Sipos2008,Ritschel2013} $1T$-Fe$_x$Ta$_{1-x}$S$_2$ (0 $\leq x \leq$ 0.05),\cite{Li2012EPL,Ang2012PRL} or $1T$-TaS$_{2-x}$Se$_x$ (0 $\leq x \leq$ 2),\cite{Liu2013,Ang2014} uniformly show the SC coexists with the NCCDW phase. In $1T$-TaS$_2$, the Mott CCDW order almost completely gaps the Fermi-surface (FS) due to strong electron-electron interaction so that it probably leaves no states for the superconducting condensate. As it melts into the NCCDW phase with the formation of CDW domains, the SC resides in.\cite{Li2012EPL,Liu2013} Therefore, we further selected $1T$-TaSSe as the parent material, with $T_c^{onset} =$ 3.8 K and $T_{NCCDW}$ = 392 K, to establish the relationship of SC and CDW domain. Besides the hydrostatic pressure method, the element doping can be effective to influence the CDW superstructure and the SC state.

In this letter, a series of Fe-substituted $1T$-Fe$_x$Ta$_{1-x}$SSe (0 $\leq x \leq$ 0.1) single crystals was fabricated via the chemical-vapor-transport (CVT) method. The XRD patterns were obtained on a Philips X'pert PRO diffractometer with Cu $K_\alpha$ radiation ($\lambda = $ 1.5418 {\AA}). The average stoichiometry was determined using X-ray energy dispersive spectroscopy (EDS) with a scanning electron microscopy (SEM). The EDS results indicate that the actual concentration $x$ is close to the nominal one. The resistivity ($\rho$) and magnetic susceptibility ($\chi$) were measured in Quantum Design PPMS-9 and MPMS-5.

The first principles calculations were performed employing the norm-conserving pseudopotential plane wave method based on DFT using the ABINIT code.\cite{gonze2009,gonze2002,gonze_brief_2005} The local-density approximation (LDA) parameterized by Perdew and Wang (PW92)\cite{PW92} was used to exchange-correlation functional in our calculations, since it can predict the accurate lattice parameters compared with our experimental data. The plane-wave cut-off energy was set to be 800 eV. We use a $\sqrt{13}\times\sqrt{13}$ superlattice in Ta-plane to model the CCDW superstructure within domains.\cite{Scruby1975,Kikuchi1998} Brillouin zone (BZ) sampling was performed on a
Monkhorst-Pack (MP) $k$-point mesh\cite{monkhorst1976} of $6\times6\times12$ for the structural relaxation, and an $8\times8\times16$ MP grid was chosen for the electronic structure calculations. The Fermi-surface was calculated on a dense $k$-point mesh grid of $16\times16\times24$.

\begin{figure}
\includegraphics[width=0.9\columnwidth]{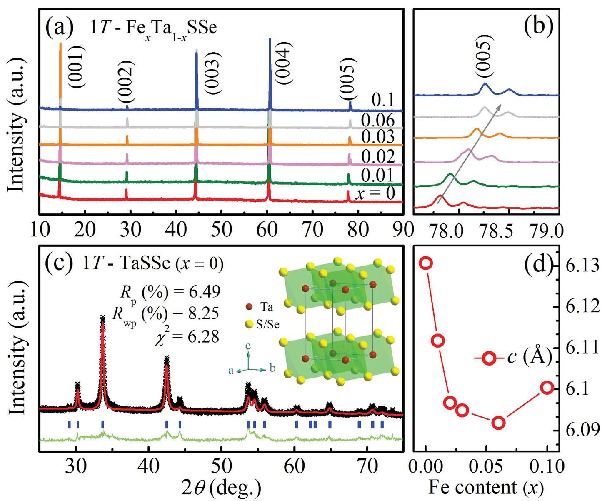}
\caption{(a) Single-crystal XRD patterns of $1T$-Fe$_x$Ta$_{1-x}$SSe (0 $\leq x \leq$ 0.1). (b) The enlargement of peak (005). (c) Powder XRD pattern with Rietveld refinement for $1T$-TaSSe ($x=$ 0). Inset shows the crystal structure of $1T$-TaSSe with randomly occupied S and Se atoms. (d) The evolution of lattice parameter $c$.}
\label{Fig1}
\end{figure}

Figure~\ref{Fig1}(a) shows the XRD patterns of $1T$-Fe$_x$Ta$_{1-x}$SSe (0 $\leq x \leq$ 0.1) single crystals at room temperature, in which only (00$l$) reflections were observed, indicating the crystallographic $c$ axis is perpendicular to the crystal plane. With increasing $x$, the diffraction peaks slightly shift to higher angle degree, which is illustrated by the enlargement of peak (005) shown in Fig.~\ref{Fig1}(b), indicating small decrease of layer space. It is mainly due to the smaller radius of Fe than that of Ta atom. Figure~\ref{Fig1}(c) shows the refinement result of powder $1T$-TaSSe. All the reflections can be indexed in the $P$-3m1 space group and the lattice parameters are obtained to be $a(=b)=$ 3.4173(8){\AA} and $c=$ 6.1466(4){\AA}. The crystal structure shows the Ta-plane is sandwiched by randomly occupied chalcogen (S and Se) layers coordinating the central Ta atom in an octahedral arrangement. Figure~\ref{Fig1}(d) shows the evolution of lattice parameter $c$, which decreases from 6.13 {\AA} for $x =$ 0 to 6.09 {\AA} for $x =$ 0.06. A slight upturn of $c$ at $x =$ 0.1 ($c =$ 6.10 {\AA}) is caused by the reduction of actual S/Se ratio. The calculated maximum variation of $c$ to be only 0.6\% ($\delta= (c_{x = 0.06} - c_{x = 0})/c_{x = 0}$), indicating a weak lattice effect.

\begin{figure}
\includegraphics[width=0.9\columnwidth]{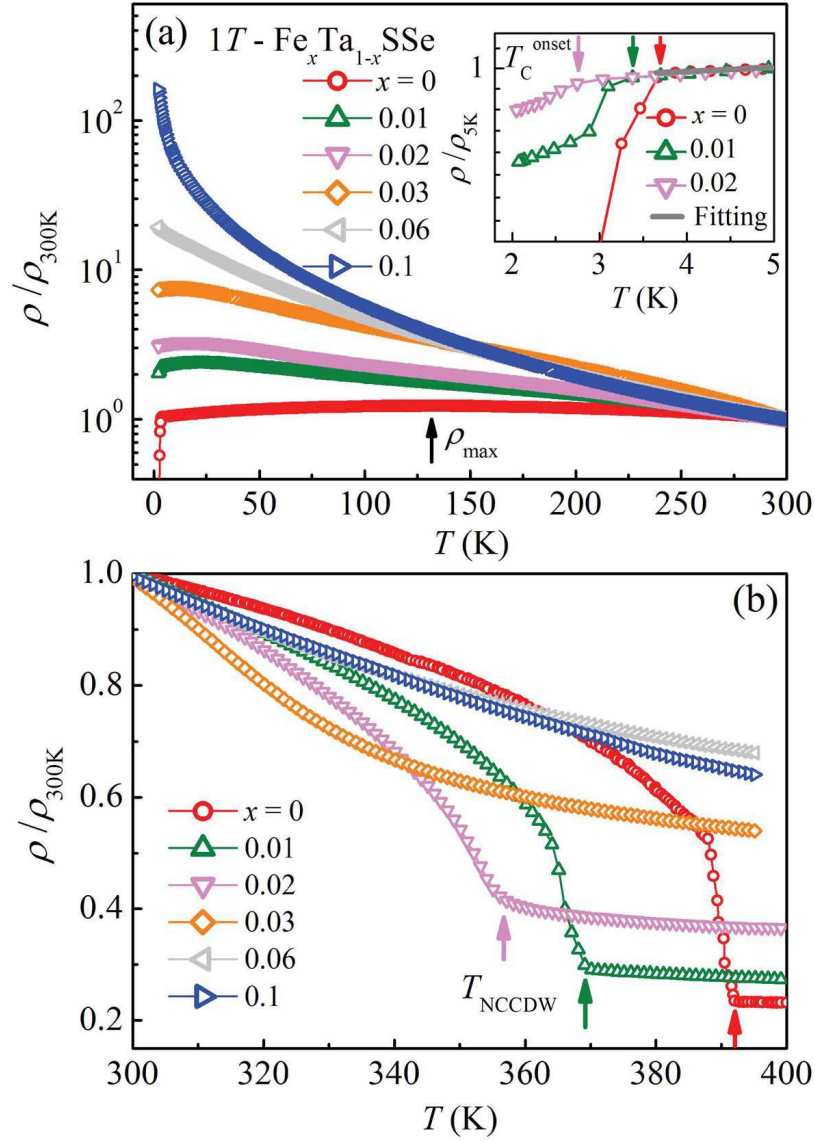}
\caption{(a) Temperature dependence of in-plane resistivity ratio $\rho/\rho_{300K}$ ($T\leq$ 300 K). The inset magnifies the superconducting region. (b) $\rho/\rho_{300K}$ ($T\geq$ 300 K).}
\label{Fig2}
\end{figure}

Figure~\ref{Fig2} shows the temperature-dependent in-plane resistivity ratio $\rho/\rho_{300K}$. Compared with the case of $1T$-TaS$_2$, in $1T$-TaSSe the Mott CCDW phase is entirely melted into the NCCDW phase.\cite{Liu2013} At low temperature, $1T$-TaSSe exhibits SC with the optimal $T_c^{onset} =$ 3.8 K, accompanied with a resistivity maximum point at $T=$ 133.8 K. With increasing $x$, the $T_c^{onset}$ decreases monotonously and could not be observed for $x =$ 0.03 within the measurement limitation $T \geq$ 2 K, as shown in the inset of Fig.~\ref{Fig2}(a). Additionally, we fitted the normal-state resistivity from 3.8 K to 12 K for $1T$-TaSSe using the power-law equation $\rho(T) = \rho_0 + AT^\alpha$, where $\rho_0$ is residual resistivity and $A$ is a constant. The fitting parameter $\alpha$ value of 0.92(4) shows nearly $T$-linear dependence, indicating the electron-phonon scattering dominates. Furthermore, the temperature of $\rho_{max}$ decreases to $T=$ 10.8 K for $x =$ 0.03, which is probably due to the enhanced Fe-substitution disorder. As $x \geq$ 0.06, the resistivity rapidly increases with decreasing temperature in the low temperature range, which is possibly ascribed to the disorder-induced Anderson localization (AL),\cite{MottPRSLS1975} as suggested previously in Fe-doped $1T$-TaS$_2$.\cite{Li2012EPL,Salvo1976PRL}

\begin{figure}
\includegraphics[width=0.9\columnwidth]{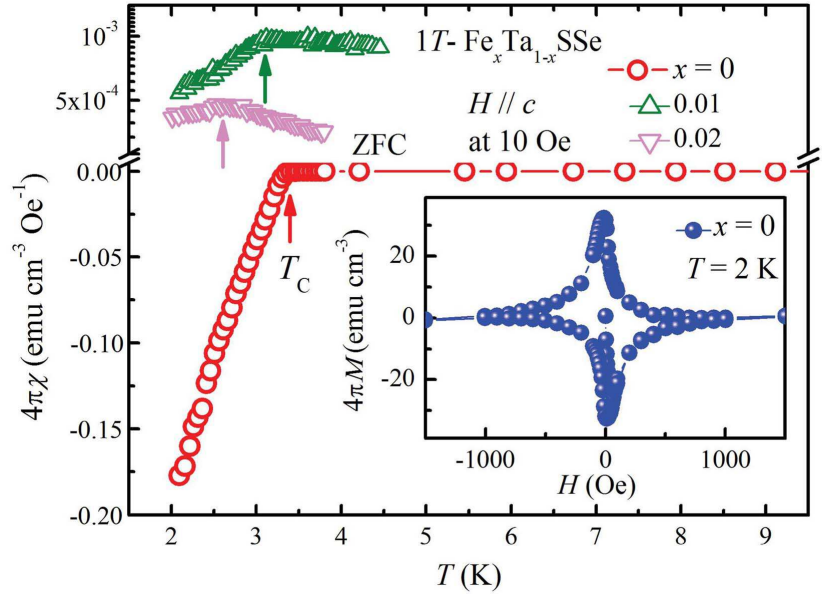}
\caption{Temperature dependence of magnetic susceptibility ($4\pi\chi$) for $1T$-Fe$_x$Ta$_{1-x}$SSe ($x =$ 0, 0.01, 0.02) with $H=$ 10 Oe. Inset: the $M(H)$ curve for $x =$ 0 at 2 K.}
\label{Fig3}
\end{figure}

In Fig.~\ref{Fig2}(b), an obvious resistivity drop was observed in $1T$-TaSSe at $T_{NCCDW} =$ 392 K. As reported in $1T$-TaS$_{2-x}$Se$_x$ (0 $\leq x \leq$ 2),\cite{Liu2013} the 392 K NCCDW in $1T$-TaSSe is enhanced from the 355 K NCCDW in $1T$-TaS$_2$. In the NCCDW phase, several tens of David-stars organize into hexagonal domains. Variable temperature STM study indicates that the CDWs are approximately commensurate within the domains and there is a phase shift between domains.\cite{Wu1991} With increasing $x$, the NCCDW transition is gradually suppressed, and disappears at $x=0.03$. Most significantly, the level ($x=0.03$) is the same as that at which the SC state disappears. It suggests that the relationship between SC and the NCCDW phase is coexistent in $1T$-Fe$_x$Ta$_{1-x}$SSe system. In addition, we measured the magnetic susceptibility ($4\pi\chi$) to establish the SC state of the samples ($x =$ 0, 0.01, 0.02) with zero-field-cooling mode (ZFC), as shown in Fig.~\ref{Fig3}. Undoubtedly, the obvious diamagnetic signal occurs at $\sim$ 3.5 K for $x =$ 0, demonstrating its SC state. Although it is not diamagnetic for $x =$ 0.01 and 0.02, an obvious transition is observed, in consistent with the resistivity data. It is mainly due to the small superconducting volume fraction. The inset of Fig.~\ref{Fig3} shows the $M(H)$ curve for $x =$ 0 at 2 K, indicating a typical type-II behavior.

\begin{figure*}
\includegraphics[width=0.95\columnwidth]{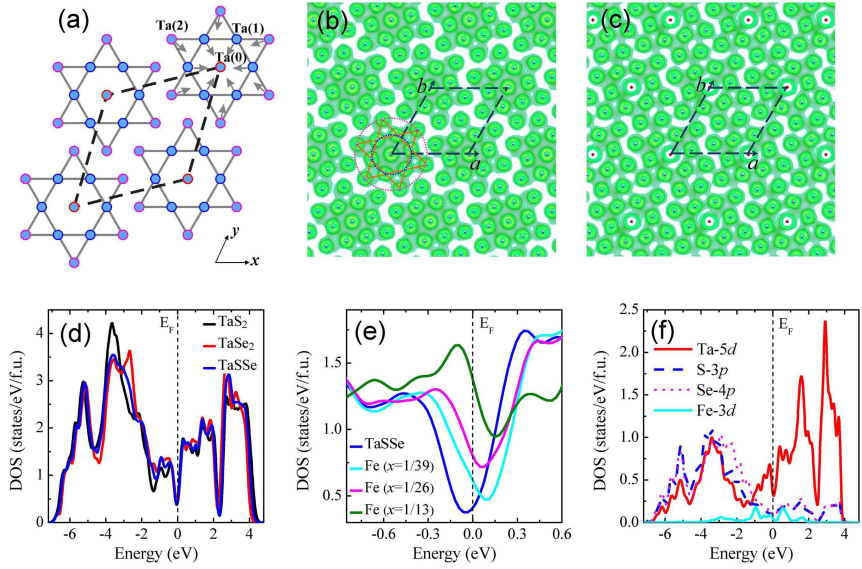}
\caption{(a) Ta basal plane with lattice distortion caused by the $\sqrt{13}\times\sqrt{13}$ superstructure. In the \textquotedblleft Stars-of-David\textquotedblright clusters, Ta(0), Ta(1), and Ta(2) are respectively corresponding to the center atoms, the first, and second rings of Ta atoms. (b) Electron density contour map in Ta-plane of $1T$-TaSSe with CCDW superlattice. (c) Electron density contour map in Ta-plane of Fe-substituted $1T$-Fe$_x$Ta$_{1-x}$SSe ($x=1/26$). The red points denote the Fe atoms. (d) DOS for $1T$-TaS$_2$ (black), $1T$-TaSe$_2$ (red), and $1T$-TaSSe (blue) in the CCDW superstructure. (e) DOS near the Fermi level for $1T$-TaSSe (blue) and $1T$-Fe$_x$Ta$_{1-x}$SSe with $x=1/39$ (cyan), $x=1/26$ (magenta), and $x=1/13$ (olive). (f) Orbital-projected DOS for $1T$-Fe$_x$Ta$_{1-x}$SSe with $x=1/26$.}
\label{Fig4}
\end{figure*}

In order to understand the Fe-substitution effect, we subsequently performed the DFT calculations. It is well known for $1T$-TaS$_2$ and $1T$-TaSe$_2$, in CCDW phase the Ta displacements form a star-like cluster (the so-called \textquotedblleft Stars-of-David\textquotedblright), which is interlocked by forming a triangular superstructure with a $\sqrt{13}\times\sqrt{13}$ periodicity (see Fig.~\ref{Fig4}(a)).\cite{Scruby1975,Kikuchi1998} In the NCCDW phase of $1T$-TaSSe, the CDWs within the domain is commensurate, which allows us to adopt such a CCDW superstructure as $1T$-TaSSe initial structure model. The initial structure was fully relaxed by taking into the lattice parameter and atom position optimizations, and the optimized structure was used for property calculations. For $1T$-TaSSe, as shown in Fig.~\ref{Fig4}(b), the CCDW-induced reconstruction of lattice is reproduced as the cases in $1T$-TaS$_2$ and $1T$-TaSe$_2$ systems. In the CCDW superlattice, Ta has three inequivalent sites including the center, the first, and second rings of \textquotedblleft Stars-of-David\textquotedblright cluster (named by Ta(0), Ta(1), and Ta(2) in Fig.~\ref{Fig4}(a)). Our calculation shows that the structure with Fe locating at Ta(0) site has the lowest energy, indicating the structure stability. The substitution that Fe prefers to locate at the center position is not surprising, since it can be easy to attain the force balance of atoms. We found that the Fe-substitution can effectively inhibit the CCDW superstructure, which is different from the case of S/Se-substitution. As shown in Fig.~\ref{Fig4}(c), the CCDW cluster has almost disappeared for $x=$ 1/26.

Figure~\ref{Fig4}(d) shows the calculated density of states (DOS) of $1T$-TaS$_2$, $1T$-TaSe$_2$, and $1T$-TaSSe. The overall character of DOS is similar and the pseudo-gap appears at the Fermi level ($E_F$), which is consistent with the previous experimental and theoretical reports in $1T$-TaS$_2$ and $1T$-TaSe$_2$.\cite{Kim1996PRB,Pillo2000PRB,Bovet2004PRB} Such pseudo-gap is due to the localization of Ta-$5d$ electrons in CCDW-reconstructed structure. In order to simulate the low Fe-doping level in our experiment, we use the maximum $\sqrt{13}\times\sqrt{13}\times3$ superlattice which corresponds to the minimum $x=1/39$. The pseudo-gap is gradually suppressed with increasing $x$ and the dip in DOS moves to higher energy (see Fig.~\ref{Fig4}(e)). The increase of DOS at $E_F$ ($N(E_F)$) indicates the enhancement of metallic conduction. In Fig.~\ref{Fig4}(f), the projected DOS shows the $N(E_F)$ is dominated by the Ta-$5d$ states and the little S-$3p$ and Se-$4p$ states. The delocalization of Ta-$5d$ electrons is attributed to the suppression of CCDW superstructure due to the Fe-substitution.

\begin{figure}
\includegraphics[width=0.95\columnwidth]{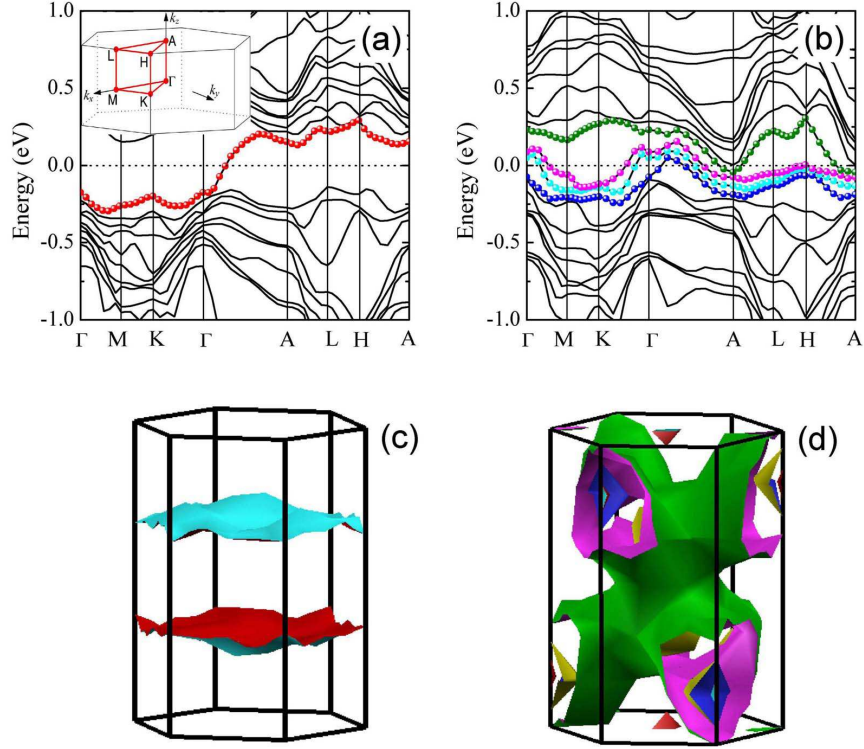}
\caption{Band structures of (a) $1T$-TaSSe ($x=0$) and (b) $1T$-Fe$_x$Ta$_{1-x}$SSe ($x=1/13$) in the $\sqrt{13}\times\sqrt{13}\times1$ superlattice. The bands crossing the Fermi level are plotted with colored points. The top inset in (a) shows the high symmetry directions within the Brillouin zone. The calculated FS of (c) $1T$-TaSSe ($x=0$) and (d) $1T$-Fe$_x$Ta$_{1-x}$SSe ($x=1/13$).}
\label{Fig5}
\end{figure}

Since Fe-substitution effectively destroys the CCDW superstructure, one can expect the band structure and FS can be well modified in $1T$-Fe$_x$Ta$_{1-x}$SSe.  As shown in Fig.~\ref{Fig5}(a), for $1T$-TaSSe, the Ta-5$d$ state becomes localized in the in-plane directions, which results in a localized uppermost band
along $\Gamma$-$M$-$K$-$\Gamma$ at about -0.3 eV. The band at $E_F$ disperses only along $\Gamma$-$A$ direction, indicating the existence of quasi-one-dimensional FS (see Fig.~\ref{Fig5}(c)), which is similar to the case of $1T$-TaS$_2$.\cite{Bovet2003PRB,Li2012EPL} However, for $1T$-Fe$_x$Ta$_{1-x}$SSe, the band structure changes especially around $E_F$. For comparison, here we show the band structure of $1T$-Fe$_x$Ta$_{1-x}$SSe with $x=1/13$ in Fig.~\ref{Fig5}(b). The four bands across the $E_F$ and have large dispersion along $\Gamma$-$M$-$K$-$\Gamma$ directions, which is related to the delocalization of Ta-$5d$ electrons in the basal plane. The hole and electron pockets compose the FS (see Fig.~\ref{Fig5} (d)) with the three-dimensional character.

Our present calculations show the Fe-substitution can effectively suppress the CDW order, consistent with our experimental observation. Once the CDW order is fully suppressed, the SC synchronously disappears. The electron-phonon (e-p) interaction may play an important role in CDW formation of $1T$-TaSSe, as argued by Liu \emph{et al}. in $1T$-TaS$_2$.\cite{Liu2009PRB} The weakening of electron-phonon interaction associated with the deformation of CDW domains could be one of the reasons for the disappearance of SC. In addition, we must point out that in the theoretical calculations we did not consider the disorder induced by Fe-substitution, which exists in real samples. The high-level substituted $1T$-Fe$_x$Ta$_{1-x}$SSe ($x\geq0.06$) shows the possible disorder-induced AL\cite{MottPRSLS1975} is understandable, as the case in Fe-substituted $1T$-TaS$_2$.\cite{Li2012EPL,Salvo1976PRL}

\begin{figure}
\includegraphics[width=0.9\columnwidth]{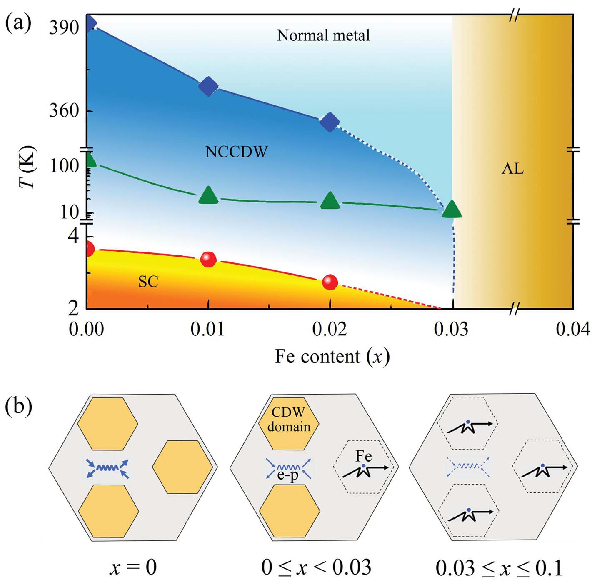}
\caption{(a)The electronic phase diagram of $1T$-Fe$_x$Ta$_{1-x}$SSe (0 $\leq x \leq$ 0.1). (b)The orange hexagons represent the CDW domains. With increasing $x$, the CDW domain is gradually destroyed, accompanied with weaker electron-phonon (e-p) interaction, and simultaneously, the Fe-substitution induced disorder scattering grows.}
\label{Fig6}
\end{figure}

Finally, we summarize the overall electronic phase diagram of $1T$-Fe$_x$Ta$_{1-x}$SSe (0 $\leq x \leq$ 0.1), as shown in Fig.~\ref{Fig6}. The NCCDW is gradually suppressed with slight Fe-substitution, more effectively than by the hydrostatic pressure method. The NCCDW and SC states are synchronously suppressed as $x\leq0.03$. With increasing $x$, it finally turns into the possible disorder-induced AL state. Figure~\ref{Fig6}(b) depicts a schematic diagram of the effect of Fe-substitution on the CDW, electron-phonon coupled SC, and disorder scattering. In $1T$-TaSSe ($x=0$), the hexagonal CDW domains form and the electron-phonon coupling is strong enough to support the CDW and SC orders. With increasing $x$, the CDW domain is gradually destroyed and at the same time, the disorder-induced scattering center emerges. The electron-phonon interaction becomes weaker, accompanying with the deformation of CDW domain, which causes the suppression of SC. As $x>0.03$, the CDW domain is almost broken and the density of scattering centers are significantly enhanced, which results in the possible AL state. The Fe-substitution at Ta-site gives a new insight that the SC and CDW domain is coexistent in $1T$-TaSSe system.

This work was supported by the National Key Basic Research under Contract No.2011CBA00111 and the National Nature Science Foundation of China through Grant No.U1232139 and the Director's Fund under Contract No.YZJJ201311 of Hefei Institutes of Physical Science, Chinese Academy of Sciences. The calculations were partially performed at the Center for Computational Science, CASHIPS.



\begin{thebibliography}{100}

\bibitem{Tranquada1995}
J. M. Tranquada, B. J. Sternlieb, J. D. Axe, Y. Nakamura, and S. Uchida, Nature {\bf375}, 561 (1995).

\bibitem{Canfield2010}
P. C. Canfield, and S. L. Bud'ko, Ann. Rev. Cond. Matt. Phys. {\bf1}, 27 (2010).

\bibitem{Rossnagel2011}
K. Rossnagel, J. Phys.:Condens. Matter {\bf23}, 213001 (2011).

\bibitem{DiSalvo1976}
F. J. Disalvo, D. E. Moncton, and J. V. Waszczak, Phys. Rev. B. {\bf14}, 4321 (1976).

\bibitem{PRL2009}
A. F. Kusmartseva, B. Sipos, H. Berger, L. Forr\'{o}, and E. Tuti\v{s}, Phys. Rev. Lett. {\bf103}, 236401 (2009).

\bibitem{Gabovich1987}
A. M. Gabovich, V. A. Medvedev, D. P. Moiseev, A. A. Motuz, A. F. Prikhotko, L. V. Prokopovich, A. V. Solodukhin, L . I. Khirunenko, V. K. Shinkarenko, A. S. Shpigev, and V. E. Yacnmenev, Fizika Nizkikh Temperatur {\bf13}, 844 (1987).

\bibitem{Chang2012}
J. Chang, E. Blackburn, A. T. Holmes, N. B. Christensen, J. Larsen, J. Mesot, R. Liang, D. A. Bonn, W. N. Hardy, A. Watenphul, M. v. Zimmermann, E. M. Forgan, and S. M . Havden, Nat. Phys. {\bf8}, 871 (2012).

\bibitem{Kartsovnik2004}
M. V. Kartsovnik, D. Andres, W. Biberacher, P. D. Grigoriev, E. A. Schuberth, and H. M\"{u}ller, J. Phys. IV France {\bf114}, 191 (2004).

\bibitem{Briggs1981}
A. Briggs, P. Monceau, M. N. Regueiro, M. Ribault, and J. Richard, J. Phys. (Paris) {\bf42}, 1453 (1981).

\bibitem{Joe2013}
Y. I. Joe, X. M. Chen, P. Ghaemi, K. D. Finkelstein, G. A. de la Pe\~{n}a, Y. Gan, J. C. T. Lee, S. Yuan, J. Geck, G. J. MacDougall, T. C. Chiang, S. L. Cooper, E. Fradkin, and P. Abbamonte, Nature Phys. {\bf10}, 421 (2014).

\bibitem{Wilson1975}
J. A. Wilson, F. J. Di Salvo, and S. Mahajan, Adv. Phys. {\bf24}, 117 (1975).

\bibitem{Fazekas1979}
P. Fazekas, and E. Tosatti, Philos. Mag. B {\bf39}, 229 (1979).

\bibitem{Scruby1975}
C. B. Scruby, P. M. Williams, and G. S. Parry, Philos. Mag. {\bf31}, 255 (1975).

\bibitem{Whangbo1992}
M. H. Whangbo, and E. Canadell, J. Am. Chem. Soc. {\bf114}, 9587 (1992).

\bibitem{Spijkerman1997}
A. Spijkerman, J. L. de Boer, A. Meetsma, G. A. Wiegers, and S. van Smaalen, Phys. Rev. B {\bf56}, 13757 (1997).

\bibitem{Sipos2008}
B. Sipos, A. F. Kusmartseva, A. Akrap, H. Berger, L. Forr\'{o}, and E. Tuti\v{s}, Nat. Mater. {\bf7}, 960 (2008).

\bibitem{Ritschel2013}
T. Ritschel, J. Trinckauf, G. Garbarino, M. Hanfland, M. v. Zimmermann, H. Berger, B. B\"{u}chner, and J. Geck, Phys. Rev. B {\bf87}, 125135 (2013).

\bibitem{Li2012EPL}
L. J. Li, W. J. Lu, X. D. Zhu, L. S. Ling, Z. Qu, and Y. P. Sun, Europhys. Lett. {\bf97}, 67005 (2012).

\bibitem{Ang2012PRL}
R. Ang, Y. Tanaka, E. Ieki, K. Nakayama, T. Sato, L. J. Li, W. J. Lu, Y. P. Sun, and T. Takahashi, Phys. Rev. Lett. {\bf109}, 176403 (2012).

\bibitem{Liu2013}
Y. Liu, R. Ang, W. J. Lu, W. H. Song, L. J. Li, and Y. P. Sun, Appl. Phys. Lett. {\bf102}, 192602 (2013).

\bibitem{Ang2014}
R. Ang, Y. Miyata, E. Ieki, K. Nakayama, T. Sato, Y. Liu, W. J. Lu, Y. P. Sun, and T. Takahashi, Phys. Rev. B {\bf88}, 115145 (2013).

\bibitem{gonze2009} X. Gonze, B. Amadon, P.-M. Anglade, J.-M. Beuken, F. Bottin, P. Boulanger, F. Bruneval, D. Caliste, R. Caracas, M. C\^{o}t\'{e}, T. Deutsch, L. Genovese, P. Ghosez, M. Giantomassi, S. Goedecker, D. R. Hamann, P. Hermet, F. Jollet, G. Jomard, S. Leroux, M. Mancini, S. Mazevet, M. J. T. Oliveira, G. Onida, Y. Pouillon, T. Rangel, G.-M. Rignanese, D. Sangalli, R. Shaltaf, M. Torrent, M. J. Verstraete, G. Zerah, and J. W. Zwanziger, Comput. Phys. Comm. {\bf180}, 2582 (2009).

\bibitem{gonze2002} X. Gonze, J.-M. Beuken, R. Caracas,
F. Detraux, M. Fuchs, G.-M. Rignanese, L. Sindic, M. Verstraete, G.
Zerah, F. Jollet, M. Torrent, A. Roy, M. Mikami, P. Ghosez, J.-Y.
Raty, and D. C. Allan, Comput. Mater. Sci. {\bf25}, 478 (2002).

\bibitem{gonze_brief_2005} X. Gonze, G.-M. Rignanese, M. Verstraete, J.-M. Beuken, Y. Pouillon, R. Caracas, F. Jollet, M. Torrent, G. Zerah, M. Mikami, P. Ghosez, M. Veithen, J.-Y. Raty, V. Olevano, F. Bruneval, L. Reining, R. W. Godby, G. Onida, D. R. Hamann, and D. C. Allan, Z. Kristallogr. {\bf220}, 558 (2005).

\bibitem{PW92}
J. P. Perdew, J. A. Chevary, S. H. Vosko, K. A. Jackson, M. R. Pederson, D. J. Singh, and C. Fiolhais, Phys. Rev. B {\bf 46}, 6671 (1992).

\bibitem{Kikuchi1998}
A. Kikuchi, Surf. Sci. {\bf 397}, 288 (1998).

\bibitem{monkhorst1976} H. J. Monkhorst, and J. D. Pack, Phys. Rev. B {\bf13}, 5188 (1976).

\bibitem{MottPRSLS1975}
N. Mott, M. Pepper, S. Pollitt, R. H. Wallis, and C. J. Adkins, Proc. Roy. Soc. Lond. A {\bf345}, 169 (1975).

\bibitem{Salvo1976PRL}
F. J. Di Salvo, J. A. Wilson, and J. V. Waszczak, Phys. Rev. Lett.
{\bf36}, 885 (1976).

\bibitem{Wu1991}
X. L. Wu, and C. M. Lieber, J. Vac. Sci. Technol. B. {\bf9}, 1044 (1991).

\bibitem{Kim1996PRB}
J.-J. Kim, I. Ekvall, and H. Olin, Phys. Rev. B {\bf54}, 2244 (1996).

\bibitem{Pillo2000PRB}
T. Pillo, J. Hayoz, H. Berger, R. Fasel, L. Schlapbach, and P. Aebi, Phys. Rev. B {\bf62}, 4277 (2000).

\bibitem{Bovet2004PRB}
M. Bovet, D. Popovi\'{c}, F. Clerc, C. Koitzsch, U. Probst, E. Bucher, H. Berger, D. Naumovi\'{c}, and P. Aebi, Phys. Rev. B {\bf69}, 125117 (2004).

\bibitem{Bovet2003PRB}
M. Bovet, S. van Smaalen, H. Berger, R. Gaal, L. Forr\'{o}, L. Schlapbach, and P. Aebi, Phys. Rev. B {\bf67}, 125105 (2003).

\bibitem{Liu2009PRB}
 A. Y. Liu, Phys. Rev. B {\bf79}, 220515 (2009).
\end{thebibliography}
\end{document}